\RequirePackage{lineno} 

\documentclass[twocolumn,showpacs,preprintnumbers,amsmath,amssymb,nofootinbib]{revtex4}

\usepackage{graphicx}% Include figure files
\usepackage{dcolumn}% Align table columns on decimal point
\usepackage{bm}% bold math

\usepackage{epstopdf}

\def\bea{\begin{eqnarray}}
\def\eea{\end{eqnarray}}

\def\pp{\mbox{$p$-$p$}}
\def\auau{\mbox{Au-Au}}

\def\pbpb{\mbox{Pb-Pb}}
\def\aa{\mbox{A-A}}

\def\ep{\mbox{$e$-$p$}}
\def\ee{\mbox{$e^+$-$e^-$}}
\def\ppbar{\mbox{$p$-$\bar p$}}
\def\qqbar{\mbox{$q$-$\bar q$}}
\def\pt{$p_t$}
\def\yt{$y_t$}

\begin{document} 

\setpagewiselinenumbers
\modulolinenumbers[5]
\linenumbers

\preprint{version 0.5}

\title{
Universal parametrization of jet production based on parton and fragment rapidities
}

\author{Thomas A.\ Trainor}\affiliation{CENPA 354290, University of Washington, Seattle, WA 98195}

%%%%%%%%%%%%%%%%%%%%%%%%%%%%%%%

\date{\today}

\begin{abstract}

As the signature manifestation of QCD in high energy nuclear collisions jet production provides essential tests of that theory. But event-wise jet reconstruction can be complex and susceptible to measurement bias. And QCD theory in the form of Monte Carlo models of elementary collisions can also be complex and difficult to test. Therefore, it may be beneficial to construct a simple static model of jet production in p-p collisions to facilitate data comparisons and model tests.
QCD is a logarithmic theory featuring  variations with energy scale as $\log(s/s_0)$. Jet-related data such as parton fragmentation functions plotted on logarithmic rapidities exhibit self-similar scaling behavior which admits an accurate parametrization with only a few parameters. In this study we extend that method to construct a parametrization of jet (scattered parton) momentum spectra based on measured logarithmic jet production trends. The parametrization is established with ISR and Sp\=pS jet data and then extrapolated for comparison with Tevatron and LHC jet data.
The jet production model from the present study is also combined with a parametrization of \ppbar\ fragmentation functions to predict the minimum-bias jet fragment contribution to hadron \pt\ spectra. The prediction is compared with published \pp\ spectrum data to test the self-consistency of the model.

\end{abstract}


\maketitle

%%%%%%%%%
 \section{Introduction}

Detailed study of jet production in high energy nuclear collisions has proceeded over more than thirty years~\cite{ua2firstjets,ua1firstjets,isrfirstjets}. During that period jet-production cross sections and jet characteristics have been measured with a variety of methods over a range of beam energies across several experimental contexts~\cite{ua1,ua2jets,aleph,opal,d0jets,cdf1,cdf2,cmsjets,atlasjets1,atlasjets2}. Intercomparisons of data within a changing experimental landscape is challenging. It is conventional to compare various aspects of jet production with several QCD Monte Carlos (MCs)~\cite{pythia,herwig,herwig2} including multiple mechanisms and parameters also evolving with time. Whether a given MC describes some data or not and to what extent, we can ask whether the basic jet  data are actually sufficiently structured  to test complex QCD MC models.

As an alternative approach a universal jet production parametrization based on a few simple principles might be developed that describes most jet-production data accurately. The parametrization could then be used as a reference system for comparisons among data sets and data/model comparisons. Such a parametrization should be efficient (few parameters), easy to generate (algebraically simple), universal (describing at least a substantial fraction of the jet-related data volume accurately) and directly related to basic QCD principles.

In addition to event-wise reconstruction of single jets jet production includes manifestations in hadron single-particle spectra and multiparticle correlations. We can therefore establish a further goal: Any model or parametrization of jet production should be consistent with spectrum and correlation manifestations or be rejected. To establish a simple quantitative connection between jet production and the hadronic final state we combine a parametrization of scattered-parton fragmentation functions in \ppbar\ collisions with a parametrization of parton (jet) spectra from \pp\ collisions to describe quantitative aspects of hadron spectra.

In this study we follow a strategy previously applied to the parametrization of parton fragmentation functions based on rapidities. We transform measured jet momentum spectra to a rapidity variable $y_{max}$ relative to an offset (jet cutoff energy) and rescale both the cross sections and jet rapidities according to  the beam rapidity. After transformation the jet data fall on a single Gaussian locus that forms the basis for the parametrization.

The jet spectrum parametrization from this study is combined with a fragmentation function parametrization to predict the minimum-bias jet contribution to hadron spectra (spectrum hard component). Quantitative correspondence with data lends support to the jet production parametrization. The success of the jet-production model for \pp\ collision energies below 1 TeV and systematic deviations from the model at higher beam energies suggest that the eikonal model as a basis for jet production Monte Carlos is questionable at lower energies but may be applicable for higher jet and collision energies.

%%%%%%%%%%%%%%%%
This article is arranged as follows:
Sec.~\ref{paramprod} introduces a parametrized model for jet production and methods used to define it.
Sec.~\ref{ffprod} describes parametrizations of fragmentation functions (FFs) from \ee\ and \ppbar\ collisions that provide a basis for the jet production model and are used to predict jet fragment contributions to hadron spectra.
Sec.~\ref{dijetprod} describes the systematics of dijet production in 200 GeV \pp\ collisions that provide one basis for energy scaling of jet production.
Sec.~\ref{jetprod}  describes the unique UA1 low-energy jet spectrum data that provide another basis for jet production modeling.
Sec.~\ref{dijspec} compares a calculated mean fragment distribution derived from the jet model in  this study and \ppbar\ FFs with 200 GeV \pp\ spectrum hard component data.
Sec.~\ref{comprecent} compares the jet production model from this study with higher-energy jet data from the Tevatron and LHC.
Secs.~\ref{disc} and~\ref{summ} present Discussion and Summary.

%%%%%%%%%%%%%%%
\section{Parametrized production Model} \label{paramprod}

In high energy nuclear collisions experimental evidence~\cite{axialci,anomalous,ptscale,porter3,jetspec} and  theoretical arguments~\cite{hijing,fragevo} indicate that most jets are produced at low energies (near 3 GeV), and low-energy jets make substantial contributions to hadron production, especially in more-central \aa\ collisions. Full understanding of nuclear collisions then depends on an accurate description of jet-related hadron production. But theoretical descriptions of low-energy jet production and fragmentation to hadrons are still incomplete. As an alternative we can attempt to develop a simple phenomenological model to serve as an interface between experiment and theory.

In this study we establish a self-consistent parametrized jet spectrum model for low-energy jets/partons from jet production data.  We then combine the spectrum model with measured FFs from \ppbar\ collisions to predict dijet contributions to \pp\ hadron spectra. Jet and fragment production are described quantitatively down to low-energy limits (on scattering and fragmentation) which we determine. 

The jet production model is based on rapidities. QCD is a logarithmic theory wherein energy scaling  of the form $\log(s/s_0)$ is common, with $\sqrt{s_0} \leftrightarrow Q_0$ some characteristic energy scale. The basis for the present model is rapidities of the form $y = \ln[(E+p)/m_h] \approx \ln(2p/m_h)$ (with hadron mass $m_h \rightarrow m_\pi$ for unidentified hadrons) combined as differences that are equivalent to the form  $\log(s/s_0)$.

Parton fragmentation  to hadron jets and projectile-nucleon fragmentation (dissociation) to soft hadrons are similar processes.  When measured FFs are plotted on fragment rapidity for different parton energies the result is a self-similar ensemble. The FF ensemble can be rescaled logarithmically both horizontally and vertically to bring all FF data onto a single locus, modulo small variations in the model-function parameters over a large jet energy interval~\cite{eeprd}. The result is a remarkable compression of FF data revealing that very few underlying degrees of freedom are actually accessible to or require theoretical description. The simple two-parameter FF model is accurate at the percent level and facilitates theoretical calculations of jet fragment production~\cite{fragevo}.

We then argue by analogy that projectile nucleon fragmentation to soft hadrons should also follow a self-similar logarithmic dependence on projectile energy. We assume only that small-x partons released by projectile dissociation in p-p collisions follow a longitudinal distribution approximately flat near mid-rapidity and falling to zero near the beam rapidity. As with FF scaling, changes in projectile energy should result in distributions with fixed shape scaling logarithmically both vertically and horizontally. According to the principle of local parton-hadron duality~\cite{lphd} the final-state soft-hadron distribution should closely follow the small-$x$ parton distribution.

We then require two more elements for the jet production model: (a) the systematics of minimum-bias (MB) dijet production from small-$x$ partons in \pp\ collisions and (b) an empirical form of the jet spectrum on parton rapidity that will serve as the scalable jet spectrum shape. The first we obtain from previous determination of a two-component model for 200 GeV \pp\ collisions~\cite{ppprd}. The second is derived from a pioneering study of low-energy ``cluster'' production at the Sp\=pS~\cite{ua1}. The combination results in a universal jet production model relying on four parameters that accurately describes all jet spectrum data for $\sqrt{s} < 1$ TeV down to an observed lower limit on jet energy near $E_{jet} = 3$ GeV. We then combine the jet spectrum model with FFs derived from \ppbar\ collisions to predict the jet fragment contribution to \pp\ hadron spectra and compare model predictions to data.

%%%%%%%%%
 \section{Parton fragmentation to jets} \label{ffprod}
 
 Measured fragmentation functions (FFs) are hadron fragment distributions on momentum or energy conditional on leading-parton or jet energy.  FFs are derived from isolated (di)jets reconstructed within high-energy elementary collisions (e.g.\ \ee, \ep, \pp, \ppbar). Although the higher-momentum portions of high-energy FFs may be described by pQCD much of the distribution is not amenable to theory and must be measured. FFs are conventionally represented by quantity $D_\alpha^\beta(x|Q^2)$ where $\alpha$ and $\beta$ represent hadron and parton types, $x$ is the fragment momentum or energy fraction of jet energy $E_{jet}$ and $Q$ is the dijet energy scale.
 
 \subsection{Parton fragmentation in $\bf e^+$-$\bf e^-$ collisions} \label{eefrag}
 
Dijet production can be described in terms of parton energy scale $Q = E_\text{dijet} = 2E_\text{jet}$. We use rapidity variables $y = \ln[(E + p)/m_\pi]$ (hadron fragment with total momentum $p$) and $y_{max} \equiv \ln(Q/m_\pi)$ (leading parton with energy $Q/2$) to describe \ee\ FFs with $D(y|y_{max}) \equiv 2dn_{ch,j}/dy$, the fragment rapidity density per dijet into $4\pi$ acceptance.  Explicit factor 2 recalls that this quantity represents a dijet fragment multiplicity. The parametrization is $D(y|y_{max}) = 2 n_{ch,j}(y_{max}) \beta(u;p,q)/y_{max}$, where $\beta(u;p,q)$ is the unit-normal (on $u$) beta distribution, $u \approx y / y_{max} \in [0,1]$ is a normalized rapidity, and parameters $p$ and $q$ (specific to each quark-hadron combination) are nearly constant over the jet or parton energy interval of interest~\cite{eeprd}. Dijet total multiplicity $2 n_{ch,j}(y_{max})$ is determined from the shape of $\beta(u;p,q)$ (and thus parameters $p$ and $q$) by parton energy conservation. Some relations to conventional quantities are $D(y|y_{max}) \approx x D(x|Q^2)$ and $y_{max} - y \approx \xi_p = \log(1/x_p)$. 
 
\begin{figure}[h]
   \includegraphics[width=1.65in,height=1.6in]{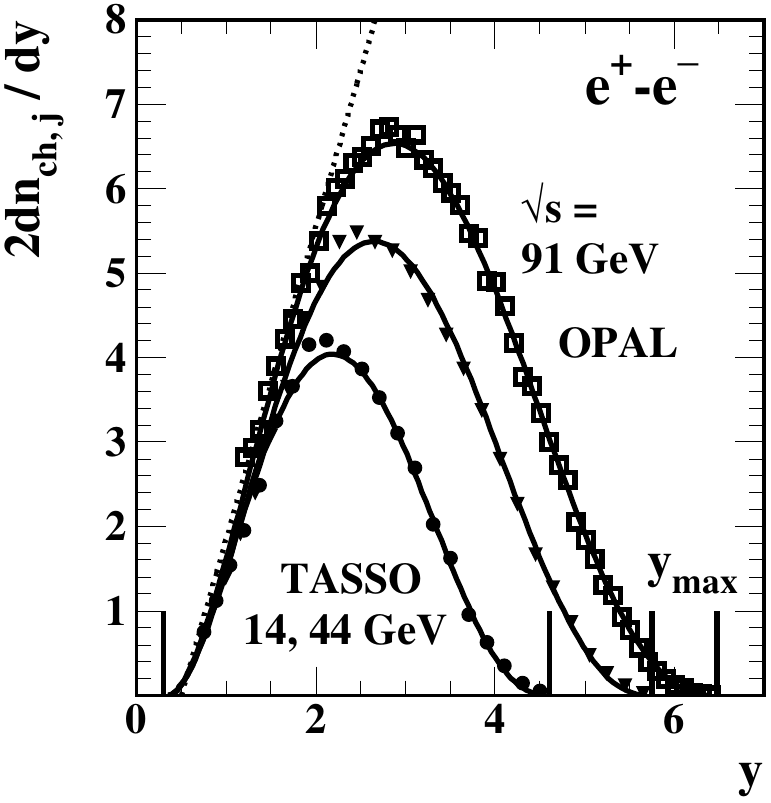}
  \includegraphics[width=1.65in,height=1.6in]{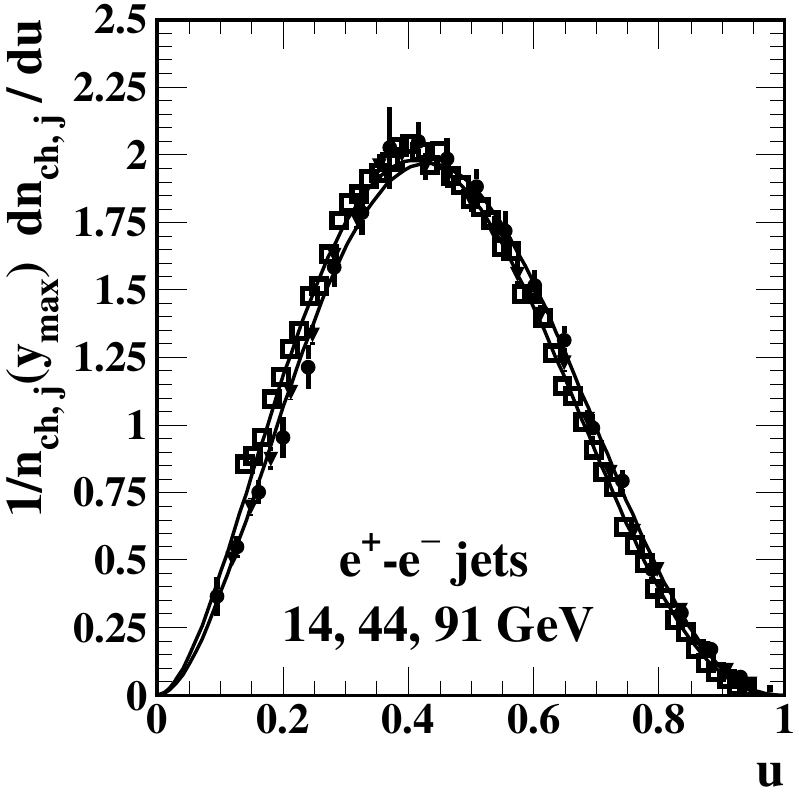}
 \caption{\label{ffs}
 Left: Fragmentation functions for three dijet energies from \ee\ collisions~\cite{tasso,opal} plotted on hadron fragment rapidity $y$ as in Ref. \cite{eeprd} showing self-similar evolution with parton rapidity $y_{max}$.
 Right: The same data rescaled to unit-normal distributions on normalized rapidity $u$. There is a barely significant evolution with parton energy. The rescaling result provides the basis for simple and accurate parametrization.
  } %fragment3c,9a
  \end{figure}

 Figure~\ref{ffs} (left panel) shows measured FFs (points) for three dijet energies derived from \ee\ collisions by TASSO~\cite{tasso} and OPAL~\cite{opal}. The data are of exceptional quality and extend down to low fragment momentum. When plotted on fragment rapidity $y$ the FFs show a self-similar evolution with parton rapidity $y_{max}$. The solid curves show the FF parametrization developed in Ref.~\cite{eeprd}.
 
 %%%%%%%%%%
 %%%%%%%%%%%%
 
 Figure~\ref{ffs} (right panel) shows the self-similar data in the left panel plotted on scaled rapidity $u = (y - y_{min}) / (y_{max} - y_{min})$ with $y_{min} \approx 0.35 $ ($p \approx 50$ MeV/c) rescaled to unit integral. The solid curves are corresponding beta distributions with parameters $p$ and $q$ nearly constant over a large jet energy interval. The simple two-parameter description is accurate to a few percent within the jet energy interval 3 GeV ($y_{max} \approx 3.75$) to 200 GeV ($y_{max} \approx 8$)~\cite{eeprd}.  FF data for light-quark and gluon jets are parametrized separately, but the parametrizations for gluon and quark jets converge near $E_{jet} = $ 3 GeV.
We find that all minimum-bias jet fragment production can be described by a few universal parameters in the context of logarithmic rapidities.

 \subsection{Comparing $\bf e^+$-$\bf e^-$ and p-\= p parton fragmentation}

Measured FFs from \ee\ and \ppbar\ or \pp\ collisions for a given dijet energy scale are quite different. Differences may arise in part from differences in event-wise jet reconstruction but also from physical differences in color connections and other QCD aspects in the two systems.

\begin{figure}[h]
   \includegraphics[width=1.65in,height=1.63in]{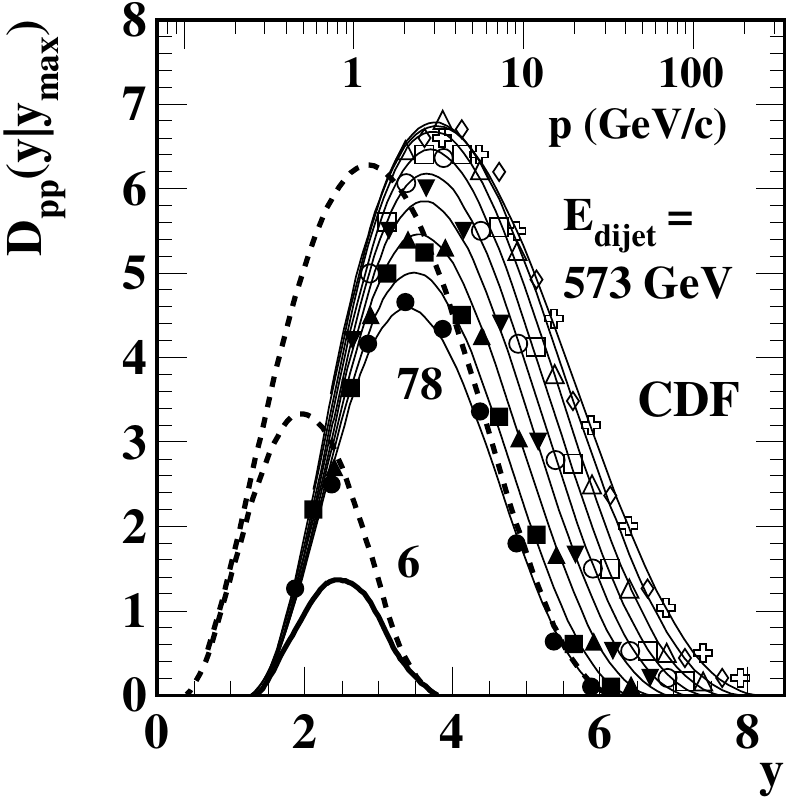}
 \includegraphics[width=1.65in,height=1.6in]{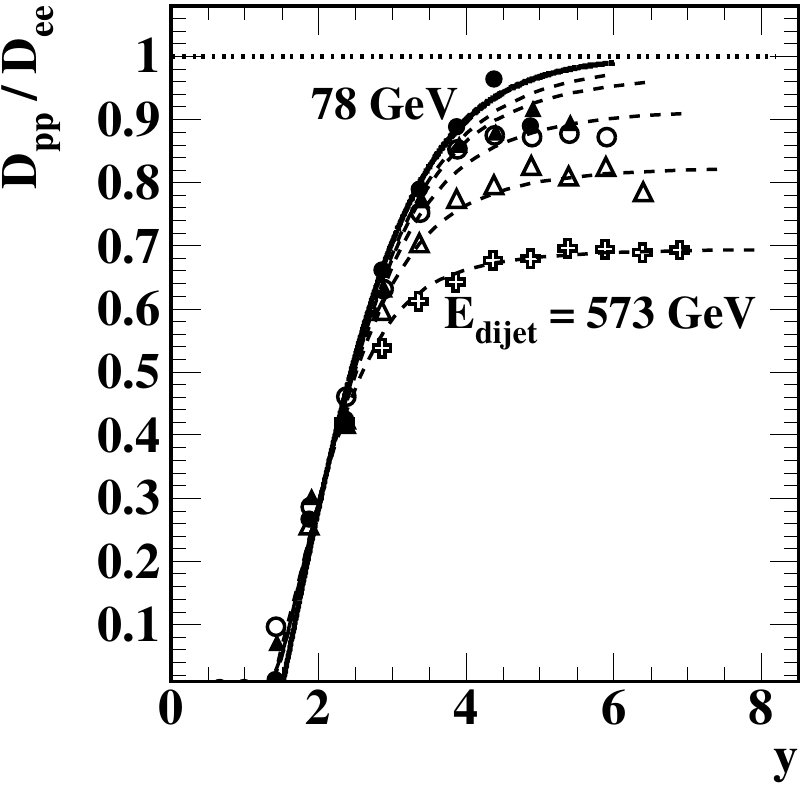}
\caption{\label{ppffs}
Left: Fragmentation functions for several dijet energies (points) from \ppbar\ collisions at 1.8 TeV~\cite{cdfff}. The solid curves represent a \ppbar\ parametrization derived from the \ee\ parametrization. The dashed curves show the \ee\ parametrization itself for two energies for comparison.
Right: The ratio of \ppbar\ FFs $D_{pp}$ to corresponding \ee\ parametrization values $D_{ee}$ vs fragment rapidity showing the systematic differences: a common strong suppression below $y = 4$ ($p \approx 4$ GeV/c) for all parton energies and a substantial reduction at larger fragment rapidities for larger parton energies.
 } %aleph17aa,aleph17b
 \end{figure}

Figure~\ref{ppffs} (left panel) shows FFs for ten dijet energies from 78 to 573 GeV inferred from 1.8 TeV \ppbar\ collisions (points) using event-wise jet reconstruction~\cite{cdfff}. Those points sample the published data distributions.  The solid curves are explained below. Comparison with the  \ee\ FF data in Fig.~\ref{ffs} (left panel) indicates that a substantial portion of dijets at lower fragment momenta may be missing from the reconstructed \ppbar\ FFs. 

%%%%%%%%%%
%%%%%%%%%%%%

Figure~\ref{ppffs} (right panel) shows the ratio of \ppbar\ FF data in the left panel to the \ee\ FF parametrization for each jet energy, revealing the systematic differences. The solid curve is $\tanh[(y-1.5)/1.7]$ which describes measured  \ppbar\ FFs relative to \ee\ FFs for jet energies below 70 GeV. The FF parametrization used in this study for \pp\ collisions (solid curves, left panel) is the \ee\ parametrization from Ref.~\cite{eeprd} modified by the tanh factor.

It is true that \ee\ FFs are observed within a full $4\pi$ acceptance whereas \ppbar\ FFs are reconstructed from a more-limited solid angle (e.g.\ pair of cones). However, we conjecture that \ee\ vs \ppbar\ differences may arise at least in part because some low-momentum part of the \ppbar\ FFs is excluded from  the mid-rapidity angular acceptance due to longitudinal transport, as discussed in Ref.~\cite{fragevo}  Sec. XIII-C. For 6 GeV dijets (lowest solid and dashed curves, left panel) \ee\ FFs give $2 n_{ch,j} \approx 5$ whereas \pp\ dijets give  $2 n_{ch,j} \approx 2$. Extrapolated to 3 GeV jets only about 40\% of the fragments in \ee\ FFs may appear in \ppbar\ FFs. The \ppbar\ FF parametrization then serves as a lower limit for comparisons in this study.

%%%%%%%%%
 \section{Dijet production in $\bf p$-$\bf p$ collisions} \label{dijetprod}

\begin{figure}[h]
  \includegraphics[width=1.65in,height=1.6in]{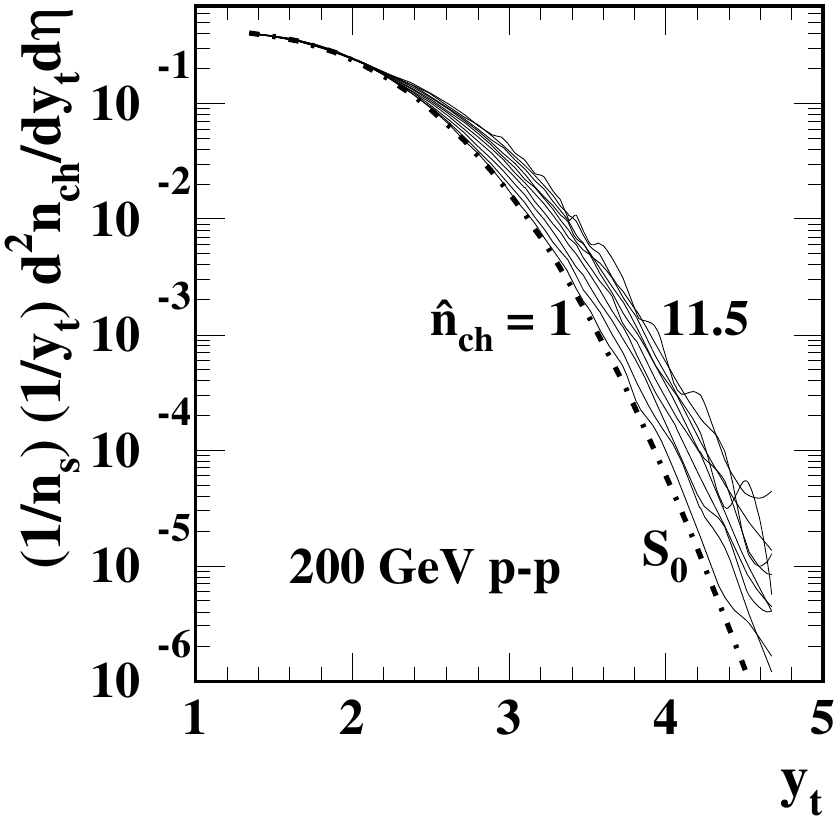}
   \includegraphics[width=1.65in,height=1.6in]{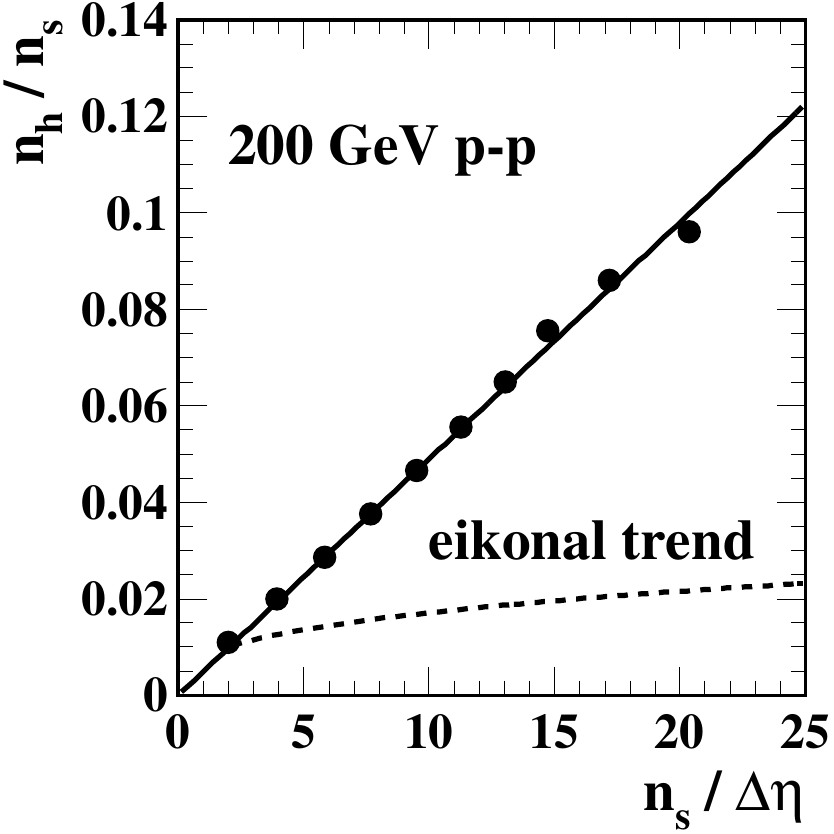}
\caption{\label{ppprd}
Left: Transverse-rapidity spectra for eleven multiplicity classes from 200 GeV NSD \pp\ collisions normalized by soft multiplicity $n_s$ within $\Delta \eta = 1$~\cite{ppprd}. Observed multiplicity $\hat n_{ch}$ is about 50\% of the corrected multiplicity $n_{ch}$. The dash-dotted curve is unit-normal soft reference $S_0$.
Right: The ratio of hard to soft multiplicity (solid dots) plotted vs the soft multiplicity density. The trend indicates that dijets scale as $n_h \propto n_s^2$ (solid line) inconsistent with the trend $\propto n_s^{4/3}$ (dashed curve) expected from the eikonal approximation. 
 } %ppcomm14d, 12cc

 \end{figure}

Figure~\ref{ppprd} (left panel) shows transverse rapidity \yt\ spectra for ten multiplicity classes from 200 GeV non-single-diffractive (NSD) \pp\ collisions normalized by soft multiplicity $n_s$ (defined below) integrated within angular acceptance $\Delta \eta = 1$ at mid-rapidity~\cite{ppprd}. Transverse rapidity for unidentified hadrons is defined as $y_t = \ln[(m_t + p_t)/m_\pi]$. The spectra are described accurately by the sum of two fixed model functions $\hat S_0(y_t)$ and $\hat H_0(y_t)$ (unit-normal soft and hard model components), the amplitudes $n_s$ (soft) and $n_h$ (hard) within some acceptance $\Delta \eta$ varying with $n_{ch} = n_s + n_h$ ($\tilde n_{ch} \approx 0.5 n_{ch}$ is the observed uncorrected multiplicity)~\cite{ppprd}.

%%%%%%%%%%
%%%%%%%%%%%%

Figure~\ref{ppprd} (right panel) shows the measured relation $n_h = \alpha n_s^2$ with $\alpha \approx 0.006$ for acceptance $\Delta \eta = 1$. Substantial evidence supports the interpretation that $n_s$ represents small-$x$ fragments from projectile proton dissociation and $n_h$ represents fragments from transverse-scattered-parton fragmentation~\cite{porter2,porter3}. That interpretation is consistent with quantitative QCD calculations derived from measured FFs and measured dijet cross sections~\cite{fragevo}. We then obtain a quantitative relation between hadron production via projectile dissociation and via scattered-parton fragmentation, with small-$x$ partons (mainly gluons) as the common element. 

Based on the argument by analogy presented above we assume that soft hadron production follows a density distribution on longitudinal rapidity or pseudorapidity varying self-similarly with beam rapidity in the form $ \Delta y_b =  y_b - y_{b0}$, where $y_{b0}$ represents an energy cutoff scale $Q_0 \approx 10$ GeV discussed below. For the self-similar system we then expect $dn_s / d\eta \propto \Delta y_b$ and $n_{s,tot} \propto (\Delta y_b)^2$ in $4\pi$ analogous to $2 n_{ch,j} \propto (y_{max} - y_{min})^2$ for FFs.

Given a spectrum hard component representing MB jets and the data in Fig.~\ref{ppprd} (right panel) we can write
\bea
dn_h/d\eta &=& f \epsilon(\Delta \eta) 2 \bar n_{ch,j} \approx 0.006 (dn_s/d\eta)^2 ,
\eea
where $f = dn_j/d\eta$ is the MB dijet density on $\eta$, $\epsilon(\Delta \eta) \in [0.5,1]$ is the fraction of a dijet that appears in $\Delta \eta$ and $2 \bar n_{ch,j}$ is the mean dijet fragment multiplicity within $4\pi$. Combined with the soft-component energy dependence above we have $f \propto (\Delta y_b)^2$ as the expected beam-energy dependence for MB dijet production at mid-rapidity.

If $n_s$ is a proxy for participant small-x partons and MB dijet production scales accurately as $n_s^2$ we can conclude  that the number of binary parton-parton collisions is $N_{bin} \propto N_{part}^2$, where $N_{part}$ represents the number of participant small-$x$ partons in a \pp\ collision. The quadratic relation implies that any combination of participant partons may result in a large-angle dijet, {\em inconsistent} with the eikonal approximation where we expect $N_{bin} \propto N_{part}^{4/3}$ or equivalently $n_h/n_s \propto n_s^{1/3}$ as shown in Fig.~\ref{ppprd} (right panel) (dashed curve).

%%%%%%%%%
 \section{ISR and S$\bf p$$\bf \bar p$S jet production} \label{jetprod}

The AFS/R807 and UA1 collaborations have separately measured jet total cross sections and differential jet \pt\ spectra down to very low jet energies for several beam energies. In the previous section we determined that MB jets (effectively the lowest-energy jets) should scale with beam rapidity as $dn_j/d\eta \propto (\Delta y_b)^2$. We now add two more assumptions: (a) differential jet spectra scale vertically in  the same way, and (b) the spectrum width on jet rapidity $y_{max}$ scales with $\Delta y_{max}$ (defined below). We rescale the measured spectra accordingly and examine the consequences.

\subsection{UA1 low-energy jet spectra}
 
\begin{figure}[h]
    \includegraphics[width=1.65in,height=1.6in]{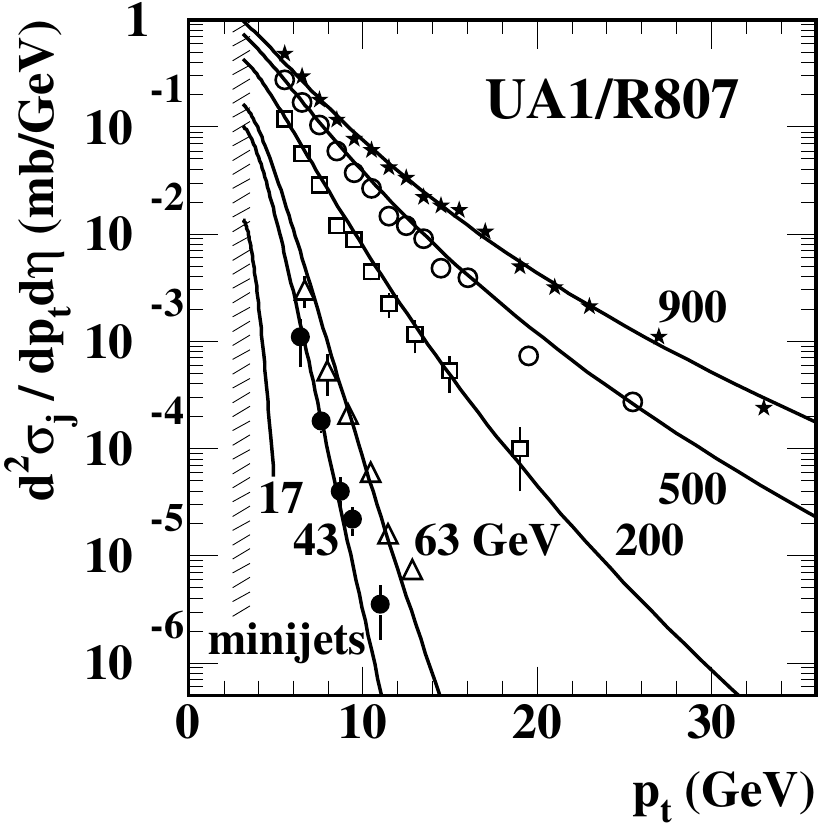}
   \includegraphics[width=1.65in,height=1.63in]{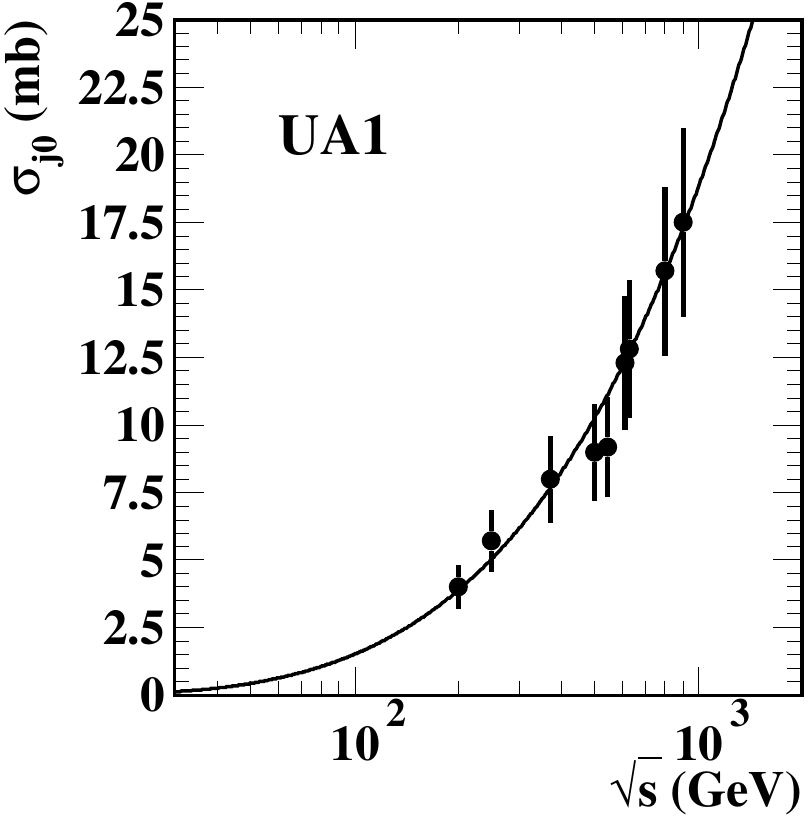}
  \caption{\label{ua1data}
  Left: Inclusive jet cross sections (points) from ISR~\cite{isrfirstjets} and Sp\=pS~\cite{ua1} collisions at five energies extending down to 5 GeV/c jet momentum. The 17 GeV curve is a model extrapolation applicable to \pbpb\ collisions at the SPS.
  Right: Jet (jet-event) total cross sections from Ref.~\cite{ua1}. The curves are described in the text.
   }    %aleph11l1, 12aa
   \end{figure}

 Figure~\ref{ua1data} (left panel) shows jet spectra for five \pp\ collision energies from the ISR (43 and 63 GeV~\cite{isrfirstjets}) and Sp\=pS (200, 500 and 900 GeV~\cite{ua1}) plotted conventionally on $p_t$. Those innovative analyses provide unique access to very low jet energies. The solid curves through the data are described below.
 
  %%%%%%%%%%
  %%%%%%%%%%%%

Figure~\ref{ua1data} (right panel) shows UA1 total cross sections for MB jet production. The curve passing through data is described below. The point-to-point deviations are small compared to the systematic-uncertainty estimates ($\pm 20$\%). However, an overall scale uncertainty factor 2 as described in Ref.~\cite{ua1} is consistent with jet-related spectrum structure described in Sec.~\ref{predict}.

In the following subsections we rescale the measured jet spectra for various energies to fall on a single model function which we determine. The result is a universal curve that can be back-transformed to describe all jet spectrum data for collision energies below 1 TeV. We can then integrate the individual spectra to obtain $d\sigma_j/d\eta$ and multiply that by an empirical expression for the effective $4\pi$ $\eta$ acceptance $\Delta \eta_{4\pi}$ to obtain the energy trend for the total cross section $\sigma_{j0}$.

\subsection{Parametrized global jet spectrum model}

The conditional jet spectrum for a given collision energy $\sqrt{s}$ is denoted by  $d^2\sigma_j /dy_{max}d\eta \equiv S_p(y_{max}|y_{b})$ with beam rapidity $y_{b}$ defined relative to pion mass as $y_b = \ln(\sqrt{s} / \text{0.14 GeV})$. Systematic analysis of available jet production data leads to a simple parametrization based on parameters $y_{b0} \equiv \ln(Q_0 / 0.14 )$ with $Q_0 \approx 10$ GeV and $y_{m0} =  \ln(2E_{cut} / 0.14 )$.  We then define $\Delta y_b = y_b - y_{b0}$ and $\Delta y_{max} = y_{b} - y_{m0}$, with normalized parton rapidity $u =  (y_{max} - y_{m0}) /\Delta y_{max}$. 

% Beam rapidity $y_b \equiv \ln(\sqrt{s} / \text{0.14 GeV})$, $y_{b0} \equiv \ln(2 \times 9 / 0.14 ) = 4.86$, $y_{m0} =  \ln(2\times 3 / 0.14 ) = 3.76$

Section~\ref{dijetprod} established that the hard-component density $dn_h/d\eta$ (and presumably dijet production $dn_j/d\eta$) in 200 GeV \pp\ collisions scales with the soft-component density as $dn_h/d\eta \propto (dn_s / d\eta)^2$. 
Given that relation and  $dn_s / d\eta \propto \Delta y_b$ at mid-rapidity
we expect dijet production at mid-rapidity to scale vertically as $dn_j/d\eta  \propto (\Delta y_b)^2$, with constant $y_{b0}$ based on a jet production cutoff near 10 GeV observed for jet-related correlations~\cite{ptedep,anomalous}. Given results in Sec.~\ref{eefrag} we also rescale jet rapidity $y_{max}$ horizontally by factor $\Delta y_{max}$ to normalized rapidity $u$. 
The data then collapse to a single locus consistent with a Gaussian if parameter $y_{m0}$ corresponds to $E_{cut} \approx 3$ GeV.

\begin{figure}[h]
     \includegraphics[width=1.65in,height=1.6in]{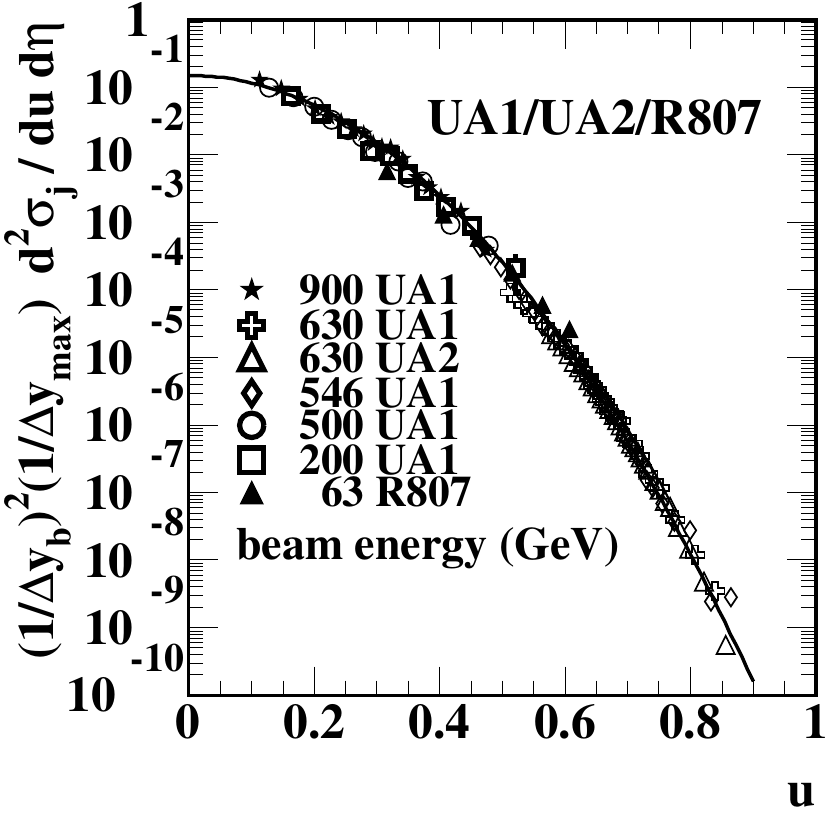}
    \includegraphics[width=1.65in,height=1.6in]{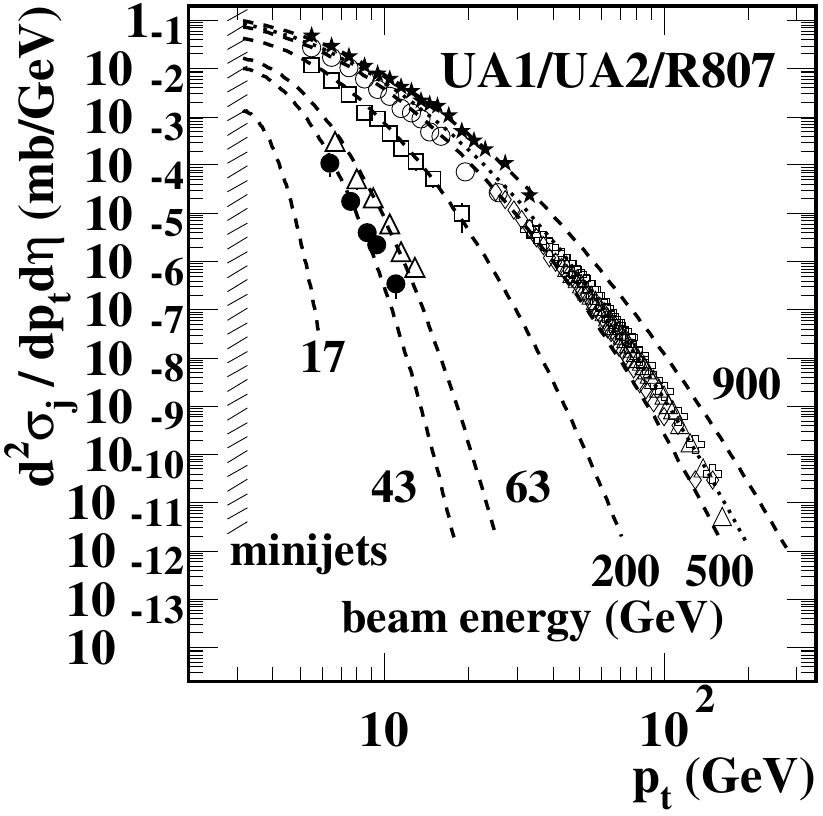}
   \caption{\label{rescale}
   Left: Rescaled jet spectra for several energies~\cite{isrfirstjets,ua1,ua2jets} plotted vs normalized fragment rapidity $u$. The data fall on a common curve $0.15 \exp(-25.5 u^2)$. 
 Right: The same data plotted in a conventional format. The dashed and dotted curves through the data are derived from the universal trend (solid curve) in the left panel as described in the text. 
 }   %aleph11l4, 11l5
    \end{figure}

Figure~\ref{rescale} (left panel) shows data from Fig.~\ref{ua1data} (left panel) with the jet spectrum (points) rescaled vertically by factor $(\Delta y_{b})^2$ and parton rapidity $y_{max}$ rescaled horizontally to $u$ by factor $\Delta y_{max}$, with $y_{m0} = 3.8$ corresponding to $E_{cut} = 3.13$ GeV. All jet data for \pp\ collision energies below 1 TeV fall on a common fitted locus $0.15 \exp(-u^2/2\sigma_u^2)$ (solid curve).
The parametrized parton spectrum conditional on beam rapidity is then 
 \bea \label{curious}
 \frac{d^2\sigma_j}{dy_{max} d\eta} &=& p_t \frac{d^2 \sigma_j}{dp_t d\eta}
\\ \nonumber
&=& 0.026 \Delta y_b^2  \frac{1}{\sqrt{2\pi \sigma^2_u}} e^{-u^2 / 2 \sigma^2_u},
 \eea
where $0.026/\sqrt{2\pi \sigma^2_u} = 0.15$ and $\sigma_u \approx 1/7$ is determined empirically from the jet  data. All jet production over nine decades is represented by parameters $y_{b0}$, $y_{m0}$, $\sigma_u$ and $\sigma_X$, the last an overall cross-section scale. Endpoints $y_{b0}$ and $y_{m0}$ are closely related by kinematic limits on charged-hadron jet production from small-$x$ partons.

%%%%%%%%%%
   %%%%%%%%%%%%

Figure~\ref{rescale} (right panel) shows the ISR and Sp\=pS cross-section data from the left panel plotted in a conventional log-log format, the curves defined by Eq.~(\ref{curious}) with beam energies noted. The dotted curve corresponds to $\sqrt{s} = 630$ GeV. All curves extend to $u = 0.9$ corresponding to partons with momentum fraction $x \approx 2/3$ where the kinematic limit of projectile-proton energy is determining. That format is used for other comparisons below.

 \subsection{Jet production energy systematics} \label{}

\begin{figure}[h]
   \includegraphics[width=1.65in]{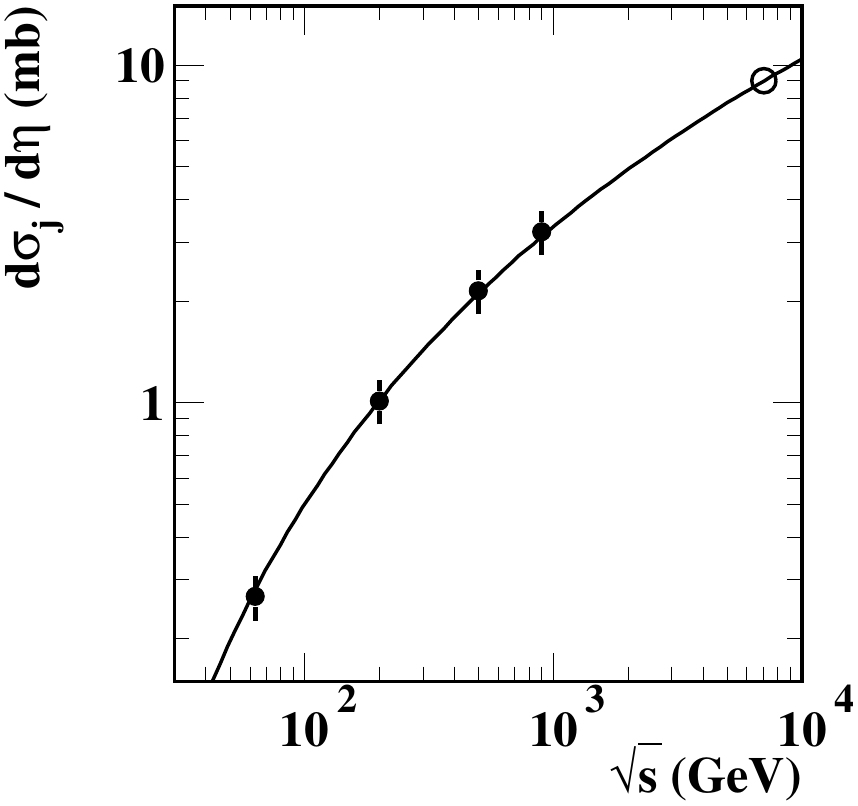}
  \includegraphics[width=1.65in]{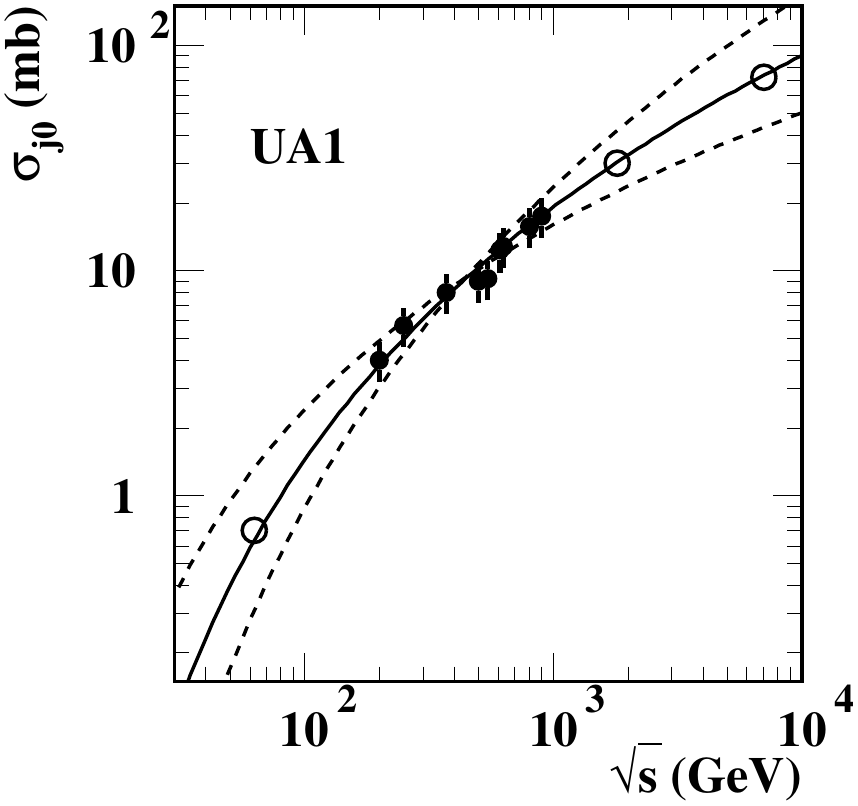}
 \caption{\label{crosssec}
 Left: Integrals of the spectra in Fig.~\ref{ua1data} (left panel) (solid points) assuming a common spectrum cutoff at $E_{cut} = 3$ GeV. The solid curve is Eq.~(\ref{sigeta}).
 Right: Jet-event total cross sections reported in Ref.~\cite{ua1} (solid points). The solid curve is  Eq.~(\ref{sigtoteq}). The dashed curves are described in the text.
  } %  aleph12b, 12a
  \end{figure}

Figure~\ref{crosssec} (left panel) shows the jet differential cross section on $\eta$ obtained by integrating Eq.~(\ref{curious})
\bea \label{sigeta}
 \frac{d\sigma_j}{d\eta} &\approx& 0.026  \Delta y_b^2 \Delta y_{max}
 \eea
which defines the solid curve. The solid points represent spectra in Fig.~\ref{ua1data} (left panel) from Refs.~\cite{isrfirstjets,ua1}. The open circle predicts the jet cross section for 7 TeV.

 %%%%%%%%%
 %%%%%%%%%%%%
 
Figure~\ref{crosssec} (right panel) shows the corresponding jet total cross section $\sigma_{j0}$.
From comparison of measured total cross sections and jet spectra in Ref.~\cite{ua1} we infer $\Delta \eta_{4\pi} \approx 1.3 \Delta y_b$.  The total cross section for jets within $4\pi$ is then 
 \bea \label{sigtoteq}
 \sigma_{j0} &=& \Delta \eta_{4\pi}  \frac{d\sigma_j}{d\eta} 
 \\ \nonumber
 &\approx &  0.034  \Delta y_b^3 \Delta y_{max}.
 \eea
plotted as the solid curve. The solid points repeated from Fig.~\ref{ua1data} (right panel) are consistent with the energy dependence of Eq.~(\ref{sigtoteq}).  The dashed curves corresponding to $0.13  \Delta y_b^2 \Delta y_{max}$ and $0.009  \Delta y_b^4 \Delta y_{max}$ provide an indication of the uncertainty in the form of Eq.~(\ref{sigtoteq}).

%%%%%%%%%%%%%%%
\section{Dijets and 200 GeV $\bf p$-$\bf p$ spectra} \label{dijspec}

In the previous section we obtained a universal jet spectrum extending down to a lower limit near 3 GeV with form consistent with general dijet production over a large collision-energy range and a dijet production cross section similarly consistent with a broad context. We now combine the jet production model from Sec.~\ref{jetprod} with FF systematics from Sec.~\ref{ffprod} to predict jet fragment contributions to hadron spectra at 200 GeV. We compare the \yt\ spectrum hard component $H(y_t)$ from 200 GeV NSD \pp\ collisions with the fragment distribution (FD) or jet-ensemble-mean FF $\bar D_u(y)$ for unidentified hadrons 

\subsection{Predicting jet-related spectrum structure} \label{predict}

The FD for unidentified hadron fragments from \pp\ collisions with beam rapidity $y_b$  is obtained by convoluting \pp\ FFs $D_u(y|y_{max})$ with parton spectrum $S_p(y_{max}|y_b)$
\bea
\bar D_u(y|y_b)  \hspace{-.02in}&=& \hspace{-.02in} \frac{\Delta \eta_{4\pi}}{\sigma_{j0}} \int \hspace{-.05in} dy_{max} D_u(y|y_{max}) S_p(y_{max}|y_b).
\eea
The per-dijet FD integrates to mean dijet multiplicity $2\bar n_{ch,j}(y_b)$.  The FD is related to the per-event spectrum hard component $y_t H(y_t,n_{ch}) \equiv d^2n_h / dy_t d\eta$ by
\bea \label{hdu}
y_t H(y_t,n_{ch}|y_b) &\approx& f(n_{ch}) \epsilon(\Delta \eta) \bar D_u(y|y_b),
\eea
where $f$ is the dijet $\eta$ density per \pp\ collision and $\epsilon(\Delta \eta)$ is the fraction of a dijet appearing in acceptance $\Delta \eta$ given the appearance of one of the jets there (the relation of $y$ to $y_t$ is discussed in Sec.~\ref{momrap}). The value $\epsilon(1) \approx 0.6$ corresponds to the $\eta$ acceptance for the analysis in Ref.~\cite{ppprd}. The value of $f$ for 200 GeV NSD \pp\ collisions is derived in the next subsection from results in Sec.~\ref{jetprod}.

\subsection{Dijet production per NSD collision at 200 GeV}

\begin{figure}[h]
  \includegraphics[width=1.65in,height=1.63in]{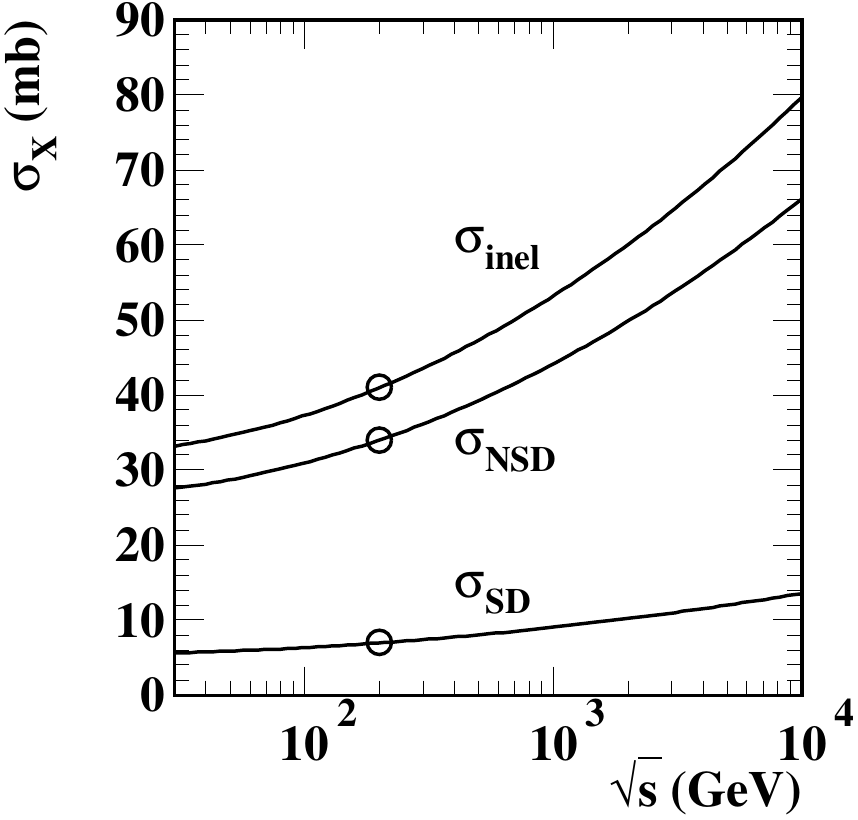}
   \includegraphics[width=1.65in,height=1.6in]{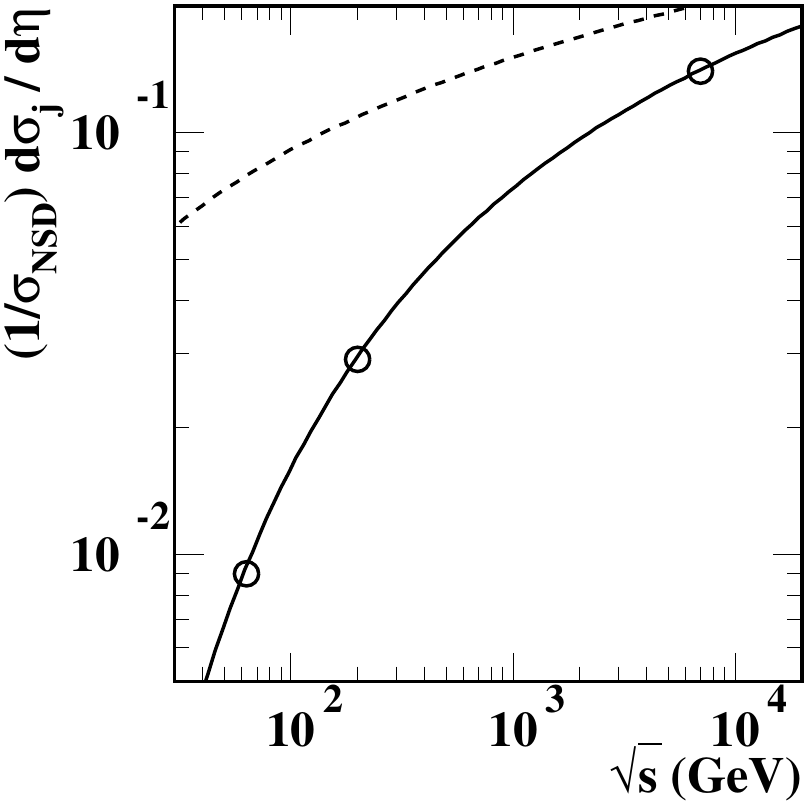}
 \caption{\label{sigma}
 Left: Curves summarizing collision-energy trends for three cross sections as reported in Ref.~\cite{alicesigma}.
 Right: The energy trend of the dijet frequency or dijet $\eta$ density per unit NSD \pp\ collision $f = (1/\sigma_{\rm NSD}) d\sigma_j /d\eta$ (solid curve) is inferred from the NSD curve in the left panel and from Eq.~(\ref{sigeta}).
  } % aleph12c, 12d
  \end{figure}

Figure~\ref{sigma} (left panel) shows parametrizations of several cross-section trends on \pp\ collision energy summarizing data shown in Ref.~\cite{alicesigma}. The inelastic cross section is described by $\sigma_\text{inel} = [32 + \Delta y_b^2]$ mb (topmost curve). The other trends are expressed as fractions $\sigma_\text{SD} = 0.17 \sigma_\text{inel}$ and $\sigma_\text{NSD} = 0.83 \sigma_\text{inel}$ (lower curves). At 200 GeV (open circles) $\sigma_\text{inel} \approx 41$ mb, $\sigma_\text{SD} \approx 7$ mb and $\sigma_\text{NSD} \approx 34$ mb. Those cross sections are used for the 200 GeV spectrum prediction below.

 %%%%%%%%%
 %%%%%%%%%%%%

Figure~\ref{sigma} (right panel) shows the predicted collision-energy trend for the $\eta$ density of dijets per NSD \pp\ collision $f_{NSD} = dn_j/d\eta = (1/\sigma_{\text NSD}) d\sigma_j / d\eta$ with value $f_{NSD} \approx 0.029$ for 200 GeV collisions  corresponding to $\sigma_{j0} \approx 4$ mb. Asymptotically, $f_{NSD}$ should increase with collision energy as $\ln(\sqrt{s} / \text{3 GeV})$ (dashed curve).

\subsection{Jet FDs vs hadron spectrum hard components} \label{fdcomp}

\begin{figure}[h]
   \includegraphics[width=1.65in,height=1.63in]{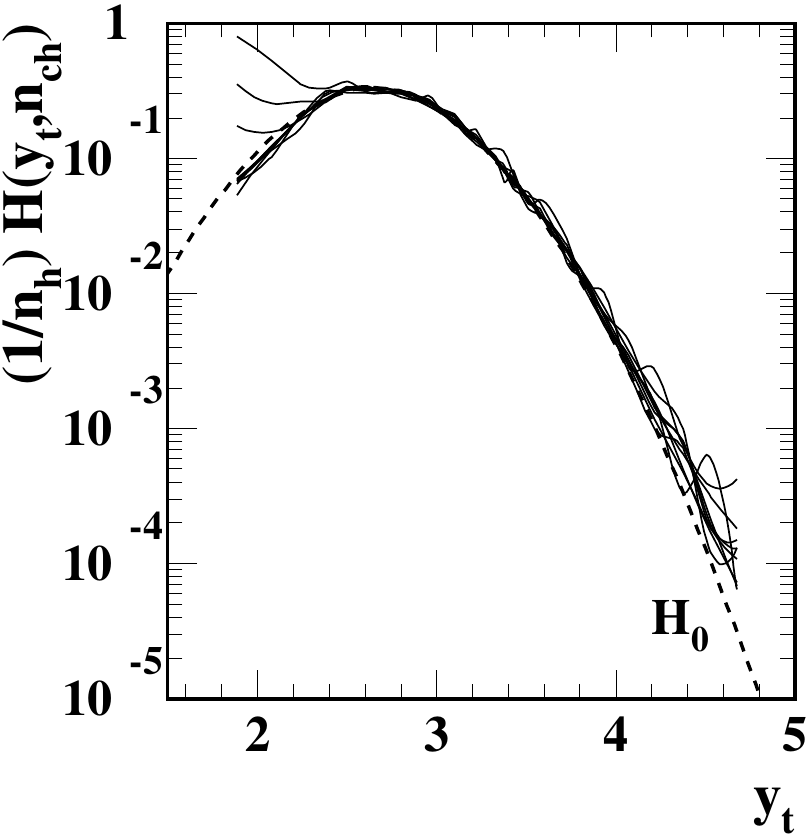}
  \includegraphics[width=1.65in,height=1.6in]{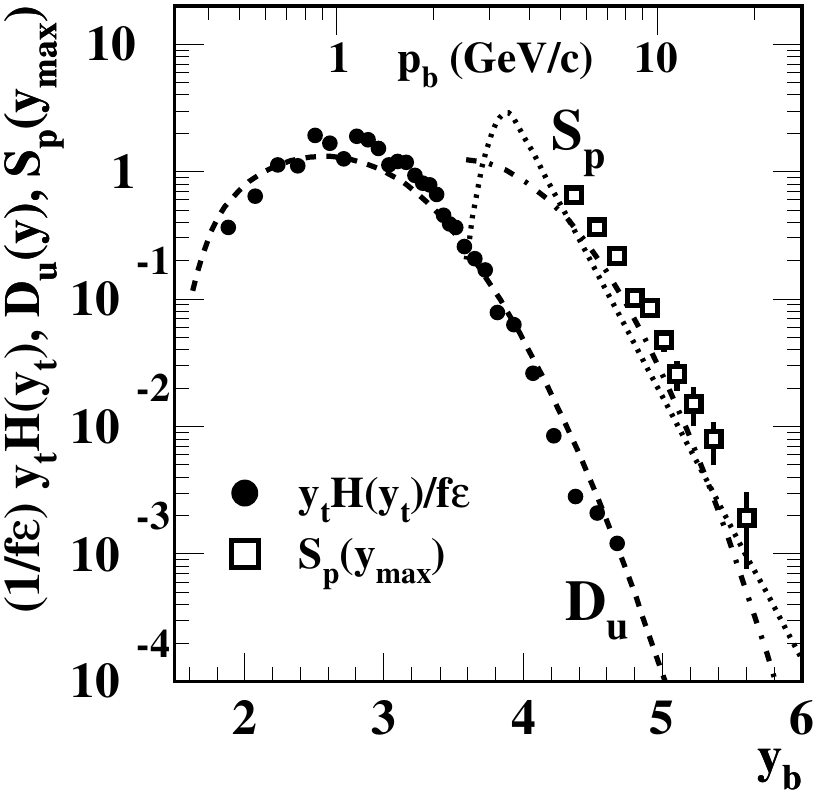}
\caption{\label{specapp}
(Color online) 
Left: Spectrum hard components $H(y_t)$ for nine multiplicity classes normalized by hard-component multiplicity $n_h$ compared to unit-normal model function $\hat H_0(y_t)$~\cite{ppprd}.
Right: The spectrum hard component for 200 GeV NSD \pp\ collisions in the form $y_t H(y_t) / f \epsilon$ (solid points) compared to calculated mean fragment distribution $\bar D_u(y)$ (dashed curve) with $y_b = y_t$, $y$ or $y_{max}$. Two spectrum models $S_p(y_{max})$ are compared with a 200 GeV jet spectrum (open squares, \cite{ua1}). The dash-dotted curve that generated $\bar D_u(y)$ is Eq.~(\ref{curious}). The dotted curve is described in the text.
 } %ppcomm12bnew, 11kppqspec
 \end{figure}

Figure~\ref{specapp} (left panel) shows unit-normal spectrum hard components in  the form $H(y_t,n_{ch}) / n_h$ from 200 NSD \pp\ collision for nine multiplicity classes (spanning the interval $n_{ch}/\Delta \eta \in [2,25]$)  corresponding to more  than a factor 100 increase in the dijet rate per \pp\ collision. The hard component is derived from the normalized spectra in Fig.~\ref{ppprd} (left panel) by subtracting fixed soft-component model $S_0(y_t)$ and dividing by an additional factor $n_h/n_s$ represented by the straight line in the right panel of that figure. The hard component has a consistent shape independent of $n_{ch}$ except for a contribution below 0.5 GeV/c ($y_t \approx 2$) for smaller $n_{ch}$. The dashed curve is the unit-normal Gaussian model $\hat H_0(y_t)$ defined in Ref.~\cite{ppprd}.

%%%%%%%%%%
%%%%%%%%%%%%

Figure~\ref{specapp} (right panel) shows hard-component data in the form $y_t H(y_t)$ from 200 GeV NSD \pp\ collisions (solid points) divided by factor $f \epsilon(\Delta \eta = 1) = 0.017$ according to Eq.~(\ref{hdu}). $\bar D_u(y)$ (dashed curve) provides a prediction for the spectrum hard component per dijet into $4\pi$ based on a parametrization of measured \pp\ FFs as in Fig.~\ref{ppffs} and a  jet spectrum model based on ISR and Sp\=pS data as discussed above. The integral of $\bar D_u(y)$ is $2\bar n_{ch,j} \approx 2.2$.
This prediction is quantitatively consistent with a previous mean fragment distribution derived from FFs and pQCD parton spectrum~\cite{fragevo} and compares well with the $y_t$ spectrum hard-component data (solid points). 

Two parton spectrum models have been employed. In Ref.~\cite{fragevo} an ad hoc power-law form was used  equivalent to $d\sigma_j/dp_t \propto 1/p_t^{5.75}$ cut off near 3 GeV ($y_{m0} \approx 3.75$) and integrating to $d\sigma_j/d\eta \approx 1.15$ mb (dotted curve). In the present study a Gaussian form defined by Eq.~(\ref{curious}) (dash-dotted curve) was inferred from ISR and Sp\=pS jet spectra. The spectrum endpoint represented by $y_{m0}$ is slightly lower (2.5 GeV) than the value (3 GeV) that best describes the UA1 200 GeV spectrum data (open squares). The Gaussian jet spectrum integrates to 0.85 mb. If the UA1 data are displaced to the left by 1 GeV or reduced by factor 2 (either adjustment is within the stated systematic uncertainties of the UA1 measurements) they fall on the parametrized parton spectrum (dash-dotted curve).

This comparison establishes that jet spectra and FFs derived from \ppbar\ collisions across several collision energies and experiments can be combined to predict the jet-related contribution to single-particle spectra down to the lowest-energy jets ($\approx 3$ GeV) and lowest-momentum hadron fragments ($\approx 0.35$ GeV/c). That result and directly-related correlation measurements provide a self-consistent quantitative description of low-energy jets and fragmentation that can be easily scaled to LHC energies as a reference for higher-energy collisions.

%%%%%%%%%
 \section{Comparison with recent jet data} \label{comprecent}

\begin{figure}[h]
 \includegraphics[width=1.65in]{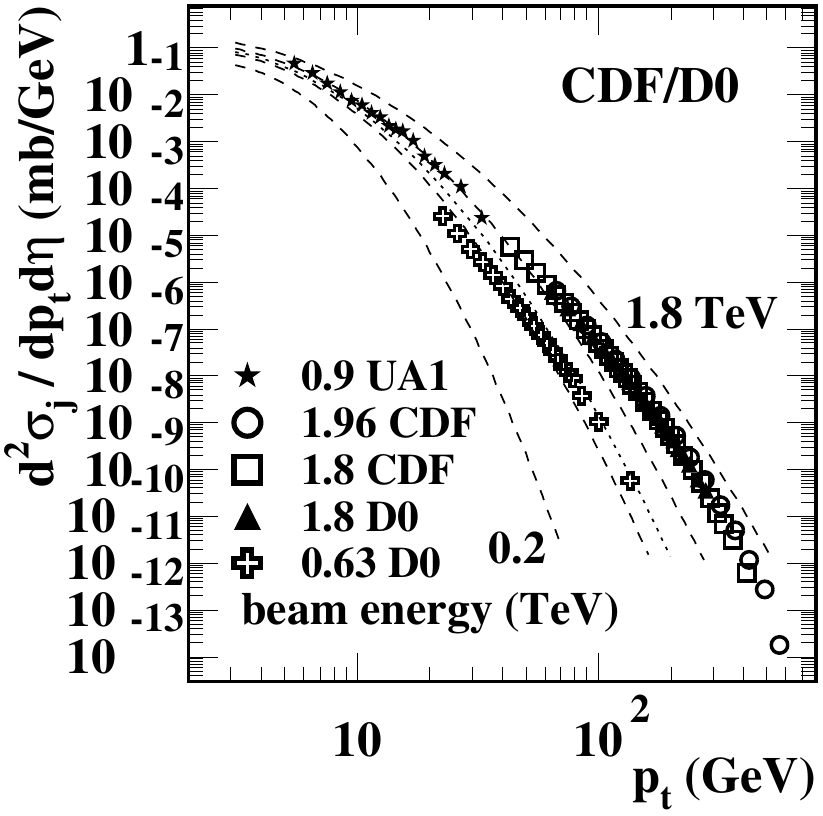}
 \includegraphics[width=1.65in]{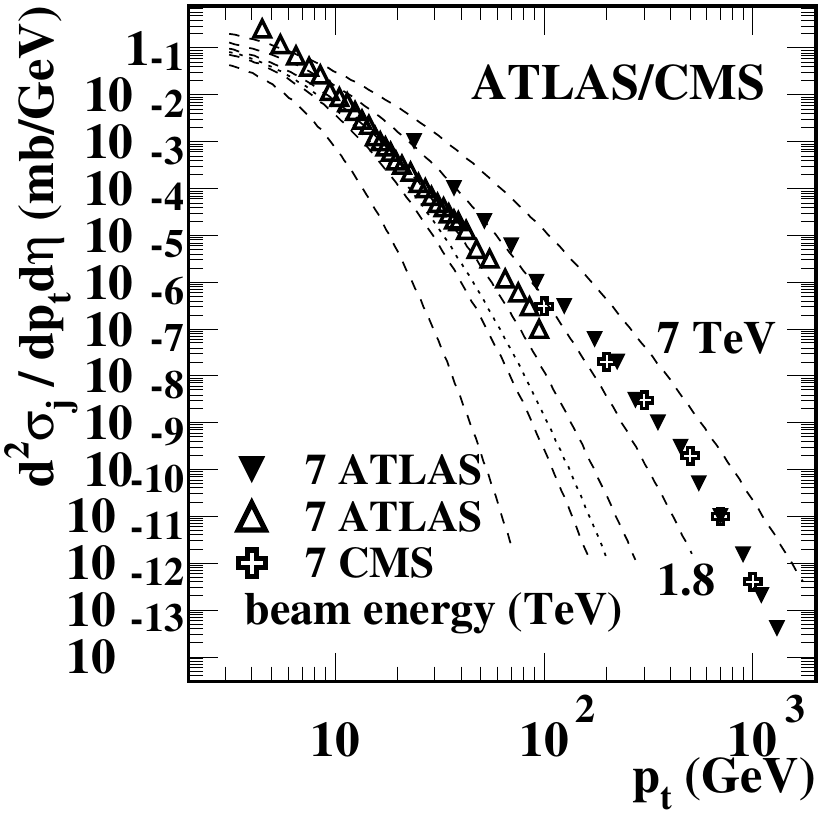}
\caption{\label{tevcomp}
Left: Tevatron jet spectra for three energies (points,~\cite{d0jets,cdf1,cdf2}) compared to the spectrum reference system (dashed and dotted curves).
Right: LHC jet spectra for 7 TeV \pp\ collisions (points,~\cite{atlasjets1,atlasjets2,cmsjets}) compared to the spectrum reference system. UA1 data for 0.9 TeV (stars) are included for comparison in the left panel.
 } % aleph11l6, 11l7
 \end{figure}

Figure~\ref{tevcomp} (left panel) shows a comparison between Tevatron jet cross sections~\cite{cdf1,cdf2,d0jets} and the reference model. The 0.63 TeV data agree with the reference (dotted curve) and Sp\=pS data for higher jet energies near 100 GeV but shift to the left of the dotted  curve for lower jet energies. The 1.8 and 1.96 TeV cross sections are a factor 9 low compared to the reference, but otherwise agree with the shape. The model curves terminate at normalized rapidity $u = 0.9$ or momentum fraction $x \approx 2/3$.

%%%%%%%%%
%%%%%%%%%%%%

Figure~\ref{tevcomp} (right panel) shows a comparison between LHC jet cross sections~\cite{atlasjets1,atlasjets2,cmsjets} and the reference system. Some of the 7 TeV data (solid triangles, open crosses) fall a factor 15 below the reference. Other 7 TeV data (open triangles, $R = 0.6$) are consistent with the UA1 0.9 TeV data (solid stars) near $E_{jet} = 50$ GeV but rise through the 7 TeV reference curve for lower jet energies. 

It should be noted that the vertical scaling according to $(\Delta y_b)^2$ adopted at lower jet and collision energies based on the non-eikonal trend $n_j \sim n_h \propto n_s^2$ may break down at larger jet and collision energies where  the corresponding transverse length scale is substantially smaller and the semiclassical eikonal approximation may be more appropriate, with consequently reduced jet cross sections. Direct comparison with hadron spectrum data in Sec.~\ref{predict} suggests that the UA1 cross sections forming the basis for the model may be a factor 1.5-2 high.

%%%%%%%%%
 \section{Discussion} \label{disc}

\subsection{What can theory Monte Carlos predict?}

The jet production model developed in this study has several features. The energy cutoffs represented by $y_{b0}$ and $y_{m0}$ are nonperturbative, determined by or consistent with previous analysis and probably both related to the observed threshold for parton fragmentation to charged hadrons near $E_{jet} = 3$ GeV. The non-eikonal $n_h \propto n_s^2$ trend used to determine the vertical scaling is empirically determined from \pp\ data. What remains is an overall cross-section scale $\sigma_X$ and the shape (approximately Gaussian) and width of the model jet spectrum on normalized rapidity $u$ that may be determined by measured proton PDFs and the parton-parton cross section determined by QCD theory. The results of this analysis, the simplicity of the jet production model, suggest that the system of PDFs on $(x,Q^2)$ may also be amenable to scaling and parametrization as shown for FFs in Fig.~\ref{ffs}.

Current \pp\ Monte Carlos such as PYTHIA~\cite{pythia} and HERWIG~\cite{herwig,herwig2} are based on the eikonal approximation and cannot therefore describe certain features of \pp\ collisions. An example can be found in Fig.~3 (top panel) of Ref.~\cite{alicempt} where standard PYTHIA (open diamonds) fails to describe $\langle p_t \rangle$ vs $n_{ch}$ data from 7 TeV \pp\ collisions, exhibiting the expected eikonal $n_{ch}^{1/3}$ trend for that case. An ad hoc color reconnection (CR) mechanism (open crosses) must be added to accommodate the data. Certain \aa\ Monte Carlos based on PYTHIA such as HIJING~\cite{hijing} and AMPT~\cite{ampt} (indirectly through HIJING on which it is based) are confronted with the same issue. Failure of HIJING to describe jet-related angular correlations in a Glauber linear superposition context is discussed in Sec.~VIII-I of Ref.~\cite{anomalous}.

\subsection{Systematic uncertainties}

The jet production model developed in  this study relies on four parameters and a Gaussian functional form. The collision-energy lower bound on dijet production that determines $y_{b0} = \ln(10/0.14) \approx 4.3$ was actually determined previously by angular correlation data and denotes an energy intercept near 10 GeV at which MB jet production appears to cease~\cite{ptedep,anomalous}. The jet production cutoff is similar to but experimentally distinct from a cutoff for nonjet quadrupole production near 13.5 GeV~\cite{davidhq}. The same lower bound applied in  this case to jet production data through factor $(\Delta y_b)^2$ results in a tight vertical correlation of data near the low-energy end of jet spectra. The same cutoff inferred from data at and below 200 GeV was used in Ref.~\cite{tomphenix} (Fig.~14, left panel) to predict the charged-hadron production  trend for \pbpb\ at 2.76 TeV. The agreement with LHC data is good. The vertical rescaling is inconsistent with the eikonal approximation applied to \pp\ collisions, as noted in Sec.~\ref{dijetprod}.

%%%%%%%%%
 \begin{figure}[h]
 \includegraphics[width=1.65in]{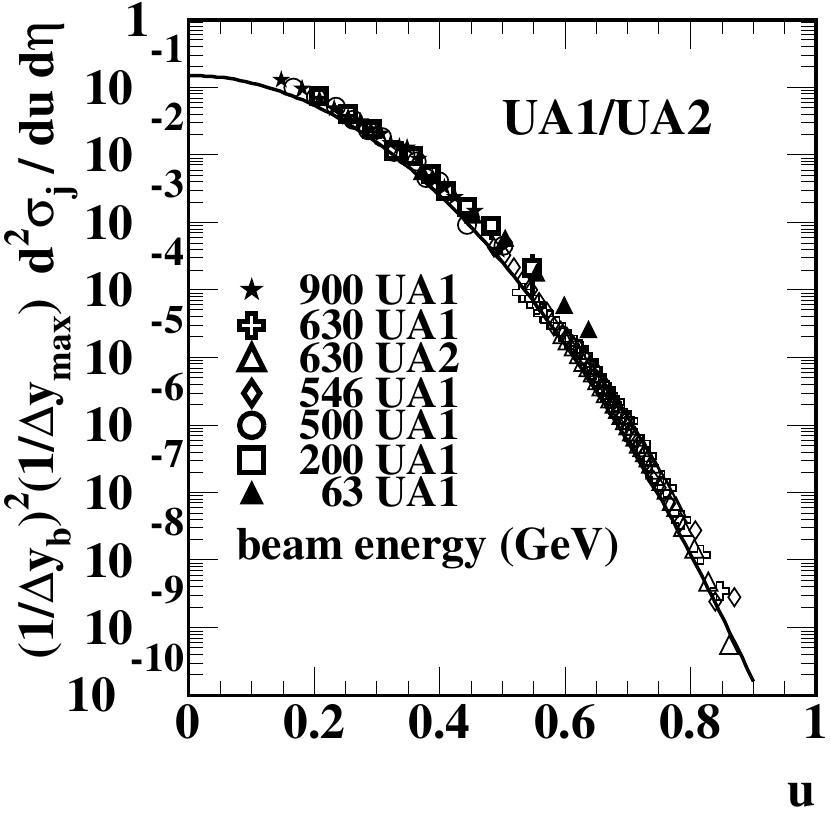}
 \includegraphics[width=1.65in]{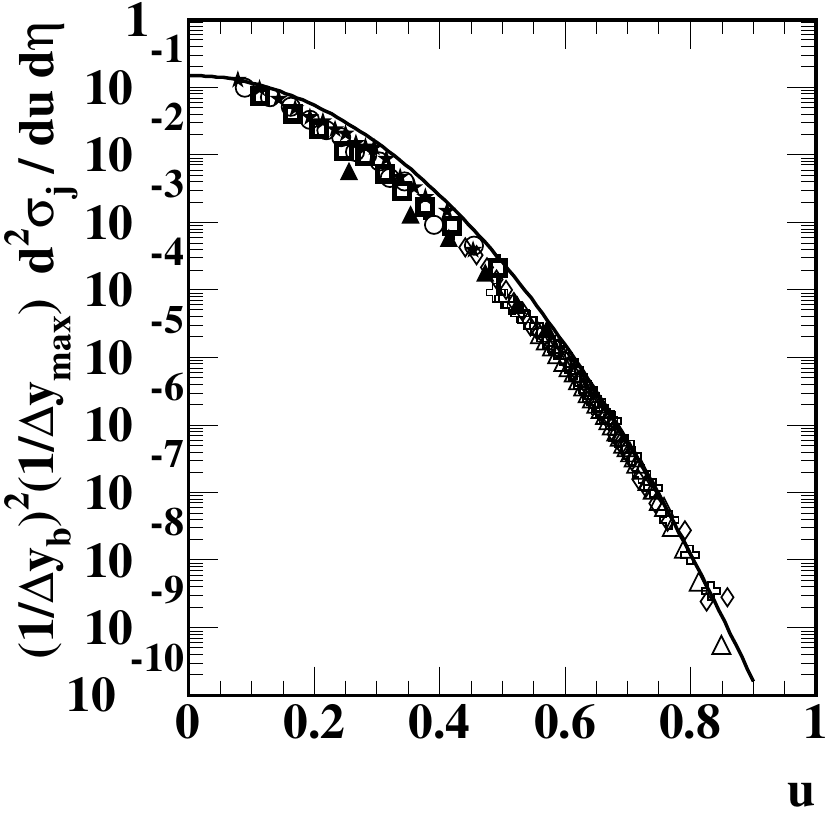}
\caption{\label{ecutsys}
Left: Data from Fig.~\ref{rescale} (left panel) with  cutoff parameter $y_{m0}$ reduced from 3.8 to 3.6 (2.6 GeV).
Right: The same with $y_{m0}$ increased from 3.8 to 4.0 (3.8 GeV).
} % aleph11l4lo, 11l4hi
 \end{figure}
%%%%%%%%%%%%

Figure~\ref{ecutsys} shows the variation of rescaled jet data as the spectrum cutoff parameter $E_{cut}$ is varied from 2.6 GeV (left panel) to 3.8 GeV (right panel). The smaller number is not excluded by the low-jet-energy UA1 data but does increase the mismatch with higher-energy jets. The larger number is ruled out by the low-jet-energy data. The scatter there is significantly increased.

Given the rescaled data the Gaussian model width is defined by $1/2\sigma_u^2 =25.5\pm1$. Variation outside that interval is excluded. The shape of the model function was tested by varying $n$ in the expression $\exp(-|u|^n/2\sigma^2_u)$. Deviations outside $n = 2.0 \pm 0.2$ are excluded by  data.

A major test of the jet production model for low-energy jets and the \pp\  FF parametrization is direct comparison with single-particle spectrum hard components from \pp\ collisions as in Sec.~\ref{fdcomp}. The shape and amplitude of the hadron spectrum hard component is consistent with a jet spectrum lower bound near 3 GeV and the Gaussian spectrum shape, but does not exclude a power-law shape. The  FF modification appearing in Fig.~\ref{ppffs} is essential to match the hadron spectrum data. Without the \ppbar\ FF cutoff the fragment distribution in Fig.~\ref{fdcomp} would greatly exceed the hadron spectrum data below 1 GeV/c. Note that the spectrum hard component is not biased by event-wise jet reconstruction and should correspond to an upper limit on jet fragment number. Nevertheless, the results are consistent with the \ppbar\ FFs. The hadron spectrum data prefer a low-energy parton spectrum $S_p$ either shifted to lower jet energy by 1 GeV or reduced in amplitude by a factor 1.5-2. Either modification is consistent with stated uncertainties in Ref.~\cite{ua1}.

It can be argued that there is a close relation between parameters $y_{b0}$ (or 10 GeV) and $y_{m0}$ (or 3 GeV) since both relate to kinematic limits on jet production. Measured jet inclusive cross sections typically fall off sharply relative to the model trends from this study beyond a momentum approximately 2/3 of the projectile energy or $x \approx 2/3$ (corresponding  to scaled rapidity $u \approx 0.9$). We also observe in hadron spectra and correlations a cutoff in parton fragmentation to charged hadrons near 3 GeV jet energy. From the jet cross-section trends we  then expect a corresponding jet production cutoff near collision energy $\sqrt{s} = 2 \times 3/2 \times 3 = 9$ GeV.
   
\subsection{Predicting jet fragment yields}

Reference~\cite{ppprd} inferred the per-event hard-component density as $dn_h/d\eta = \alpha  (dn_s/d\eta)^2  \approx 0.005 (2.5)^2 \approx 0.03$ from a two-component spectrum analysis.
The assumption was then made that $dn_h/d\eta = f 2\bar n_{ch,j}$, where $f$ is the dijet frequency within acceptance $\Delta \eta = 1$. Assuming $2\bar n_{ch,j} = 2.5$ extrapolated from CDF data yielded $f \approx 0.012$. It is interesting to reconsider that estimate based on the analysis in Ref.~\cite{fragevo} and the current study.

In Ref.~\cite{fragevo} the integrated jet total cross section was estimated to be $\sigma_{j0} \approx 2.5$ mb, the $4\pi$ $\eta$ acceptance was taken as $\Delta \eta_{4\pi} = 5$ and $\sigma_{NSD} \approx 0.87\times42 = 36.5$ mb~\cite{ua5nsd} giving $f = \sigma_{j0} / \Delta \eta_{4\pi}\sigma_{NSD} \approx 0.014$, close to the value inferred in Ref.~\cite{ppprd}. 
In the present study based on UA1 numbers $\sigma_{j0} \rightarrow 4$ mb, $ \Delta \eta_{4\pi} \rightarrow 4$ or $d\sigma_j/d\eta \approx 1$ and $\sigma_{NSD} \rightarrow 34$ mb, with $f \approx 0.029$ as in Fig.~\ref{sigma} (right panel).

The analysis in Ref.~\cite{ppprd} omitted factor $\epsilon(\Delta \eta)$ that relates dijet fragment detection in $4\pi$ as at LEP to dijet detection within a limited acceptance $\Delta \eta$, with $\epsilon(1) \approx 0.6$ relevant to that analysis. Also, the result in Fig.~\ref{ppprd} (right panel) indicates that $\alpha \approx 0.006$ is a more accurate parameter value than 0.005, leading to $dn_h / d\eta \approx 0.038$. Combining  those results we obtain $2\bar n_{ch,j} = (1/f\epsilon) dn_h/d\eta \approx 2.2$. Thus, although there were substantial changes in several factors the original assumption about mean dijet multiplicity is consistent with what is inferred directly from the spectrum data and jet production cross sections. 

In Ref.~\cite{ppprd} an extrapolated mean dijet multiplicity estimate was used to infer $f$. In  the present study a calculated value for $f$ based on high-energy jet measurements is used to infer the mean dijet multiplicity. The inferred value 2.2 (for MB \pp\ jets) is about 40\% of the expected 5.7 for LEP 3 GeV FFs and is roughly consistent with the cutoff function (solid curve) in Fig.~\ref{ppffs} (left panel).

\subsection{Momentum and rapidity components} \label{momrap}

Most of the FFs from LEP and HERA are reported in terms of fragment total momentum $p$ or momentum fraction $x_p = p/p_{jet}$ from which we infer fragment rapidity $y$. However, the ALEPH collaboration has reported fragment momenta as cylindrical components $p_z$ and $p_t$~\cite{aleph}. It is interesting to note that FFs plotted vs longitudinal (along the jet thrust axis) rapidity $y_z$ derived from $p_z$ are flat near the origin $y_z = 0$ (do not descend to zero) but fall toward zero with approach to the parton rapidity $y_{z,max}$. We conclude that the fall to zero near $y =0$ for FFs based on total momentum $p$ as in Fig.~\ref{ffs} is a result of the Jacobian between $y$ and $y_z$. The so-called ``hump-backed plateau'' is  a result of that Jacobian. The distribution on $y_z$, flat near the parton-parton center of momentum, is intuitively expected based on fragmentation of a color field (string) drawn between two recoiling partons (e.g.\ \qqbar\ pair).

In comparisons between LEP FFs based on $p$ and hadron spectra from nuclear collisions based on $p_t$ as in Sec.~\ref{fdcomp} there may also be a significant Jacobian effect. $p_t$ relative to a nuclear collision axis is approximately $p_z$ along a jet thrust axis for jets near mid rapidity. Thus, in Fig.~\ref{specapp} some disagreement between $\bar D_u(y)$ and $y_t H(y_t)$ may be expected below 1 GeV/c because of  mismatched rapidity definitions.

%%%%%%%%%%%
\section{Summary}\label{summ}

A universal parametrization of jet production in high energy nuclear collisions has been constructed and compared with inclusive jet cross sections at several energies. The model is motivated by the need to provide a fixed reference for comparisons among different jet-production data sets and between jet-related data and theory Monte Carlos, given that data analysis methods and Monte Carlos have evolved significantly over three decades.

The jet production model is based on a previous description of \ee\ fragmentation functions (FFs) derived by scaling their fragment rapidity dependence for a range of parton energies to a universal data trend described by a simple two-parameter model function. The same scaling technique is applied in  this case to jet rapidity spectra acquired for a range of projectile proton energies. The energy scaling in the model is determined by an observed non-eikonal trend of jet production in \pp\ collisions, by the measured collision-energy scaling of jet-related angular correlations in \auau\ collisions and by an observed cutoff of jet production near 3 GeV jet energy.

The model describes ISR and Sp\=pS jet data within their uncertainties for collider energies below 1 TeV and jet energies below 200 GeV. Comparisons with Tevatron and LHC jet data at higher beam and jet energies reveals smooth systematic differences that may reveal a breakdown of the model assumptions at higher energy scales and/or systematic biases in some inclusive cross sections.

The jet spectrum model evaluated at 200 GeV is combined with a parametrization of FFs derived from Tevatron \ppbar\ collisions to predict the contribution of minimum-bias jets (spectrum hard component) to the hadron \pt\ spectrum from 200 GeV NSD \pp\ collisions. The data description is good provided that the jet spectrum is shifted to lower jet energy by 1 GeV or reduced in amplitude by factor 1.5-2, either adjustment permitted by  the stated uncertainty of UA1 jet cross sections.

%%%%%%%%%%%%%%%%%%
This work was supported in part by the Office of Science of the U.S.\ DOE under grant DE-FG03-97ER41020.

%%%%%%%%%%%%%%%%%%%%%%%%%%%%

\end{document}